\documentclass[aps,prc,preprintnumbers,superscriptaddress]{revtex4}

\usepackage[dvips]{graphicx}
\usepackage{dcolumn}
\usepackage{bm}
\usepackage{longtable}
\begin{document}

% Use the \preprint command to place your local institutional report
% number in the upper righthand corner of the title page in preprint mode.
% Multiple \preprint commands are allowed.
% Use the 'preprintnumbers' class option to override journal defaults
% to display numbers if necessary
\preprint{Accepted for publication in Nature Physics (DOI: 10.1038/s41567-018-0410-7)}

%Title of paper
%\title{Competition between shell and shape \\clarified by novel magnetic moment measurement of $^{75}$Cu}
%\title{Measurement of the magnetic moment of $^{75}$Cu reveals \\the interplay between nuclear shell evolution and shape deformation}
\title{Interplay between nuclear shell evolution and shape deformation \\revealed by magnetic moment of $^{75}$Cu}

\author{Y.~Ichikawa}
%\author{Yuichi Ichikawa}
\email[]{yuichikawa@riken.jp}
\affiliation{RIKEN Nishina Center for Accelerator-Based Science, 2-1 Hirosawa, Wako, Saitama 351-0198, Japan}
\author{H.~Nishibata}
%\author{Hiroki Nishibata}
\affiliation{RIKEN Nishina Center for Accelerator-Based Science, 2-1 Hirosawa, Wako, Saitama 351-0198, Japan}
\affiliation{Department of Physics, Osaka University, 1-1 Machikaneyama, Toyonaka, Osaka 560-0034, Japan}
\author{Y.~Tsunoda}
%\author{Yusuke Tsunoda}
%\affiliation{Department of Physics, University of Tokyo, 7-3-1 Hongo, Bunkyo, Tokyo 113-0033, Japan}
\affiliation{Center for Nuclear Study, University of Tokyo, 7-3-1 Hongo, Bunkyo, Tokyo 113-0033, Japan}
\author{A.~Takamine}
%\author{Aiko Takamine}
\affiliation{RIKEN Nishina Center for Accelerator-Based Science, 2-1 Hirosawa, Wako, Saitama 351-0198, Japan}
\author{K.~Imamura}
%\author{Kei Imamura}
\affiliation{RIKEN Nishina Center for Accelerator-Based Science, 2-1 Hirosawa, Wako, Saitama 351-0198, Japan}
\affiliation{Department of Physics, Meiji University, 1-1-1 Higashi-Mita, Tama, Kawasaki,
  Kanagawa 214-8571, Japan}
\author{T.~Fujita}
%\author{Tomomi Fujita}
\affiliation{RIKEN Nishina Center for Accelerator-Based Science, 2-1 Hirosawa, Wako, Saitama 351-0198, Japan}
\affiliation{Department of Physics, Osaka University, 1-1 Machikaneyama, Toyonaka, Osaka 560-0034, Japan}
\author{T.~Sato}
%\author{Tomoya Sato}
\affiliation{RIKEN Nishina Center for Accelerator-Based Science, 2-1 Hirosawa, Wako, Saitama 351-0198, Japan}
\affiliation{Department of Physics, Tokyo Institute of Technology,
  2-12-1 Oh-okayama, Meguro, Tokyo 152-8551, Japan}
\author{S.~Momiyama}
%\author{Satoru Momiyama}
\affiliation{Department of Physics, University of Tokyo, 7-3-1 Hongo, Bunkyo, Tokyo 113-0033, Japan}
\author{Y.~Shimizu}
%\author{Yohei Shimizu}
\affiliation{RIKEN Nishina Center for Accelerator-Based Science, 2-1 Hirosawa, Wako, Saitama 351-0198, Japan}
\author{D.~S.~Ahn}
%\author{Deuksoon Ahn}
\affiliation{RIKEN Nishina Center for Accelerator-Based Science, 2-1 Hirosawa, Wako, Saitama 351-0198, Japan}
\author{K.~Asahi}
%\author{Koichiro Asahi}
\affiliation{RIKEN Nishina Center for Accelerator-Based Science, 2-1 Hirosawa, Wako, Saitama 351-0198, Japan}
\affiliation{Department of Physics, Tokyo Institute of Technology,
  2-12-1 Oh-okayama, Meguro, Tokyo 152-8551, Japan}
\author{H.~Baba}
%\author{Hidetada Baba}
\affiliation{RIKEN Nishina Center for Accelerator-Based Science, 2-1 Hirosawa, Wako, Saitama 351-0198, Japan}
\author{D.~L.~Balabanski}
%\author{Dimiter L. Balabanski}
\affiliation{RIKEN Nishina Center for Accelerator-Based Science, 2-1 Hirosawa, Wako, Saitama 351-0198, Japan}
%\affiliation{ELI-NP, IFIN-HH, 077125 M\u{a}gurele, Romania}
%\affiliation{ELI-NP, Horia Hulubei National Institute for R&D in Physics and Nuclear Engineering,
%  077125 M\u{a}gurele, Romania}
\affiliation{ELI-NP, Horia Hulubei National Institute of Physics and Nuclear Engineering, 077125 M\u{a}gurele, Romania}
\author{F.~Boulay}
%\author{Florent Boulay}
\affiliation{RIKEN Nishina Center for Accelerator-Based Science, 2-1 Hirosawa, Wako, Saitama 351-0198, Japan}
\affiliation{CEA, DAM, DIF, F-91297 Arpajon, France}
\affiliation{GANIL, CEA/DSM-CNRS/IN2P3, Bvd Henri Becquerel, F-14076 Caen, France}
\author{J.~M.~Daugas}
%\author{Jean-Michel Daugas}
\affiliation{RIKEN Nishina Center for Accelerator-Based Science, 2-1 Hirosawa, Wako, Saitama 351-0198, Japan}
\affiliation{CEA, DAM, DIF, F-91297 Arpajon, France}
\author{T.~Egami}
%\author{Tsuyoshi Egami}
\affiliation{Department of Advanced Sciences, Hosei University, 3-7-2 Kajino-cho, Koganei,
  Tokyo 184-8584, Japan}
\author{N.~Fukuda}
%\author{Naoki Fukuda}
\affiliation{RIKEN Nishina Center for Accelerator-Based Science, 2-1 Hirosawa, Wako, Saitama 351-0198, Japan}
\author{C.~Funayama}
%\author{Chikako Funayama}
\affiliation{Department of Physics, Tokyo Institute of Technology,
  2-12-1 Oh-okayama, Meguro, Tokyo 152-8551, Japan}
\author{T.~Furukawa}
%\author{Takeshi Furukawa}
\affiliation{Department of Physics, Tokyo Metropolitan University,
  1-1 Minami-Ohsawa, Hachioji, Tokyo 192-0397, Japan}
\author{G.~Georgiev}
%\author{Georgi Georgiev}
\affiliation{CSNSM, CNRS-IN2P3, Universit\'e Paris-sud, UMR8609, F-91405 Orsay-Campus, France}
\author{A.~Gladkov}
%\author{Aleksei Gladkov}
\affiliation{RIKEN Nishina Center for Accelerator-Based Science, 2-1 Hirosawa, Wako, Saitama 351-0198, Japan}
\affiliation{Department of Physics, Kyungpook National University, 80 Daehak-ro, Buk-gu, Daegu 702-701, South Korea}
\author{N.~Inabe}
%\author{Naoto Inabe}
\affiliation{RIKEN Nishina Center for Accelerator-Based Science, 2-1 Hirosawa, Wako, Saitama 351-0198, Japan}
\author{Y.~Ishibashi}
%\author{Yoko Ishibashi}
\affiliation{RIKEN Nishina Center for Accelerator-Based Science, 2-1 Hirosawa, Wako, Saitama 351-0198, Japan}
\affiliation{Department of Physics, University of Tsukuba, 1-1-1 Tennodai, Tsukuba, Ibaraki 305-8577, Japan}
\author{T.~Kawaguchi}
%\author{Takafumi Kawaguchi}
\affiliation{Department of Advanced Sciences, Hosei University, 3-7-2 Kajino-cho, Koganei,
  Tokyo 184-8584, Japan}
\author{T.~Kawamura}
%\author{Takayuki Kawamura}
\affiliation{Department of Physics, Osaka University, 1-1 Machikaneyama, Toyonaka, Osaka 560-0034, Japan}
\author{Y.~Kobayashi}
%\author{Yoshio Kobayashi}
\affiliation{Department of Informatics and Engineering, University of Electro-Communication,
  1-5-1 Chofugaoka, Chofu, Tokyo 182-8585, Japan}
\author{S.~Kojima}
%\author{Shuichiro Kojima}
\affiliation{Department of Physics, Tokyo Institute of Technology,
  2-12-1 Oh-okayama, Meguro, Tokyo 152-8551, Japan}
\author{A.~Kusoglu}
%\author{Asli Kusoglu}
\affiliation{ELI-NP, Horia Hulubei National Institute of Physics and Nuclear Engineering, 077125 M\u{a}gurele, Romania}
\affiliation{CSNSM, CNRS-IN2P3, Universit\'e Paris-sud, UMR8609, F-91405 Orsay-Campus, France}
\affiliation{Department of Physics, Faculty of Science, Istanbul University, Vezneciler/Fatih,
  34134 Istanbul, Turkey}
\author{I.~Mukul}
%\author{Ish Mukul}
\affiliation{Department of Particle Physics, Weizmann Institute of Science, Rehovot 76100, Israel}
\author{M.~Niikura}
%\author{Megumi Niikura}
\affiliation{Department of Physics, University of Tokyo, 7-3-1 Hongo, Bunkyo, Tokyo 113-0033, Japan}
\author{T.~Nishizaka}
%\author{Taishi Nishizaka}
\affiliation{Department of Advanced Sciences, Hosei University, 3-7-2 Kajino-cho, Koganei,
  Tokyo 184-8584, Japan}
\author{A.~Odahara}
%\author{Atsuko Odahara}
\affiliation{Department of Physics, Osaka University, 1-1 Machikaneyama, Toyonaka, Osaka 560-0034, Japan}
\author{Y.~Ohtomo}
%\author{Yuichi Ohtomo}
\affiliation{RIKEN Nishina Center for Accelerator-Based Science, 2-1 Hirosawa, Wako, Saitama 351-0198, Japan}
\affiliation{Department of Physics, Tokyo Institute of Technology,
  2-12-1 Oh-okayama, Meguro, Tokyo 152-8551, Japan}
\author{T.~Otsuka}
%\author{Takaharu Otsuka}
\affiliation{RIKEN Nishina Center for Accelerator-Based Science, 2-1 Hirosawa, Wako, Saitama 351-0198, Japan}
\affiliation{Center for Nuclear Study, University of Tokyo, 7-3-1 Hongo, Bunkyo, Tokyo 113-0033, Japan}
\affiliation{Department of Physics, University of Tokyo, 7-3-1 Hongo, Bunkyo, Tokyo 113-0033, Japan}
\affiliation{Instituut voor Kern- en Stralingsfysica, K. U. Leuven,
  Celestijnenlaan 200D, B-3001 Leuven, Belgium}
\author{D.~Ralet}
%\author{Damian Ralet}
\affiliation{CSNSM, CNRS-IN2P3, Universit\'e Paris-sud, UMR8609, F-91405 Orsay-Campus, France}
%\author{T.~Shimoda}
%\author{Tadashi Shimoda}
%\affiliation{Department of Physics, Osaka University, 1-1 Machikaneyama, Toyonaka, Osaka 560-0034, Japan}
\author{G.~S.~Simpson}
%\author{Gary S. Simpson}
\affiliation{LPSC, CNRS/IN2P3, Univerisit\'e Grenoble Alpes, CNRS/IN2P3, INPG,
F-38026 Grenoble, France}
\author{T.~Sumikama}
%\author{Toshiyuki Sumikama}
\affiliation{RIKEN Nishina Center for Accelerator-Based Science, 2-1 Hirosawa, Wako, Saitama 351-0198, Japan}
\author{H.~Suzuki}
%\author{Hiroshi Suzuki}
\affiliation{RIKEN Nishina Center for Accelerator-Based Science, 2-1 Hirosawa, Wako, Saitama 351-0198, Japan}
\author{H.~Takeda}
%\author{Hiroyuki Takeda}
\affiliation{RIKEN Nishina Center for Accelerator-Based Science, 2-1 Hirosawa, Wako, Saitama 351-0198, Japan}
\author{L.~C.~Tao}
%\author{Longchun Tao}
\affiliation{RIKEN Nishina Center for Accelerator-Based Science, 2-1 Hirosawa, Wako, Saitama 351-0198, Japan}
\affiliation{State Key Laboratory of Nuclear Physics and Technology, School of Physics,
  Peking University, Beijing 100871, China}
\author{Y.~Togano}
%\author{Yasuhiro Togano}
\affiliation{Department of Physics, Tokyo Institute of Technology,
  2-12-1 Oh-okayama, Meguro, Tokyo 152-8551, Japan}
\author{D.~Tominaga}
%\author{Daiki Tominaga}
\affiliation{Department of Advanced Sciences, Hosei University, 3-7-2 Kajino-cho, Koganei,
  Tokyo 184-8584, Japan}
\author{H.~Ueno}
%\author{Hideki Ueno}
\affiliation{RIKEN Nishina Center for Accelerator-Based Science, 2-1 Hirosawa, Wako, Saitama 351-0198, Japan}
\author{H.~Yamazaki}
%\author{Hiroki Yamazaki}
\affiliation{RIKEN Nishina Center for Accelerator-Based Science, 2-1 Hirosawa, Wako, Saitama 351-0198, Japan}
\author{X.~F.~Yang}
%\author{Xiaofei Yang}
\affiliation{Instituut voor Kern- en Stralingsfysica, K. U. Leuven,
  Celestijnenlaan 200D, B-3001 Leuven, Belgium}

\email[]{yuichikawa@riken.jp}
%\homepage[]{Your web page}
%\thanks{}
%\altaffiliation{}

\date{\today}

%\begin{abstract}
%\end{abstract}

%\pacs{}
%\keywords{}
%\maketitle must follow title, authors, abstract, \pacs, and \keywords
\maketitle

\clearpage

{\bf
Exotic nuclei are characterized by a number of neutrons (or protons) in excess relative to stable nuclei.
Their shell structure, which represents single-particle motion in a nucleus~\cite{mayer,jensen},
may vary due to nuclear force and excess neutrons~\cite{warner,janssens,42Si,54Ca},
in a phenomenon called shell evolution~\cite{otsuka2005}.
This effect could be counterbalanced by collective modes causing deformations of the nuclear surface~\cite{bohr-mottelson}.
Here, we study the interplay between shell evolution and shape deformation
by focusing on the magnetic moment of an isomeric state of the neutron-rich nucleus $^{75}$Cu.
We measure the magnetic moment using highly spin-controlled rare-isotope beams and achieving large spin alignment
via a two-step reaction scheme~\cite{ichikawa} that incorporates an angular-momentum-selecting nucleon removal.
By combining our experiments with numerical simulations of many-fermion correlations,
we find that the low-lying states in $^{75}$Cu are, to a large extent, of single-particle nature on top of a correlated $^{74}$Ni core.
We elucidate the crucial role of shell evolution even in the presence of the collective mode, and within the same framework,
we consider whether and how the double magicity of the $^{78}$Ni nucleus is restored,
which is also of keen interest from the perspective of nucleosynthesis in explosive stellar processes. 
}

%\clearpage

\begin{figure}[bp]
  \begin{center}
        \includegraphics[width=9.5cm]{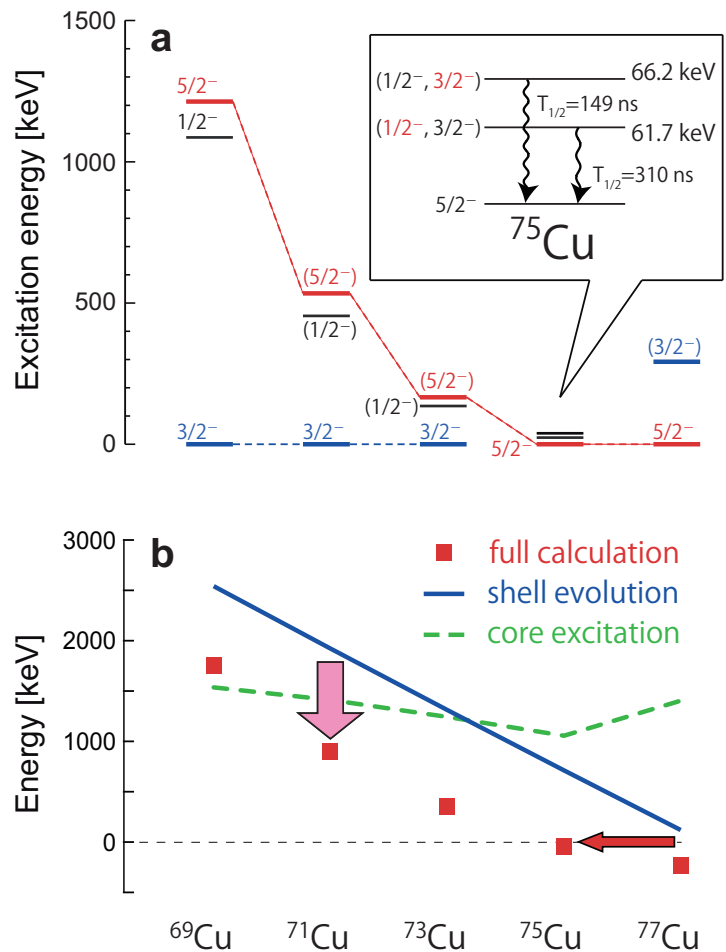}
  \end{center}
  \caption{Shell evolution in neutron-rich Cu isotopes.
    {\bf a.} Experimental systematics of energy level and spin parity for odd-$A$ Cu
    isotopes~\cite{franchoo1,flanagan,koester,sahin}.
    Red and blue bars represent the $5/2^-$ and $3/2^-$ states, respectively.
    The inset shows the low-lying isomeric states of $^{75}$Cu,
    which emit a 61.7-keV $\gamma$ ray with a half-life of $T_{1/2}=310(8)$~ns and a 66.2-keV $\gamma$ ray
    with $T_{1/2}=149(6)$~ns~\cite{petrone}.
    The spin parity values in red were identified as a result of the present experiment. 
    {\bf b.} Theoretical energy of the $5/2^-_1$ state as measured from the $3/2^-_1$ state
    calculated in three different ways with the A3DA-m Hamiltonian~\cite{tsunoda}.  
    The red squares are obtained from the full calculation with the Monte Carlo Shell Model (MCSM).  
    The blue solid line represents the single-particle energies in the na\"ive shell evolution scenario, 
    driven by the monopole interaction.  
    The green dashed line represents the effects of the core excitation, when the most relevant
    monopole interaction between the $\pi f_{5/2}$ or $\pi p_{3/2}$ orbital and the  
    $\nu g_{9/2}$ orbital is deactivated.
    The vertical pink arrow indicates schematically the effect of core excitation, 
    whereas the horizontal red arrow indicates the shift of the crossing point. 
  }
  \label{fig_cu-level}
\end{figure}

% shell evolution in Cu isotopes
The Cu isotopes, which carry 29 protons ({\it i.e.}, the atomic number is $Z=29$),
play a key role in efforts to understand the structure of exotic nuclei.  
Figure \ref{fig_cu-level}~{\bf a} displays observed low-lying energy levels of Cu isotopes with even numbers of neutrons, $N=40 \sim 48$.    
We start with a na\"ive picture in which the 29 protons consist of one ``last'' proton on top of the Ni ($Z=28$) magic core,
and in low-lying states of spin parity $I^{\pi} = 5/2^-$ or $3/2^-$, this last proton occupies the $f_{5/2}$ or $p_{3/2}$ orbital, respectively, 
with the rest of the nucleus, including the neutrons, being a spherical Ni core in its $I^{\pi} = 0^+$ ground state.

%  exp/theo levels 
Figure \ref{fig_cu-level}~{\bf a} illustrates the energy drop of the $5/2^-$ level~\cite{franchoo1} over 1~MeV
from an excited state in $^{69}$Cu down to the ground state of $^{75}$Cu~\cite{flanagan},
implying the possibility of a crossing between the proton $f_{5/2}$ and $p_{3/2}$ orbitals.  
This behaviour is recognized as one of the most ideal examples of shell evolution~\cite{otsuka2005,otsuka2010}.
Figure~\ref{fig_cu-level}~{\bf b} further shows 
the calculated energy differences between the $5/2^-$ and $3/2^-$ states based on the single-particle model mentioned above (blue solid line),
exhibiting a lowering similar to the experimental one.
This lowering occurs due to the monopole component of the nuclear forces, particularly the tensor force, 
as more neutrons are added to the $g_{9/2}$ orbit~\cite{otsuka2010}.
Figure~\ref{fig_cu-level}~{\bf b} also displays the results of a full calculation (red squares) beyond the single-particle model. 
The trends appear to be similar among the two sets of results in panel {\bf b} and the corresponding energy difference obtained from panel {\bf a}. 
Although the single-particle picture thus appears well representative, 
certain collective modes must be admixed, as hinted by the differences between the two calculations.
Note that collective modes here mean excitations of the Ni core, such as surface oscillations.
Therefore, to pin down how the shell evolves and how it competes with the collective modes,
it is necessary to consider the wave function.

% magnetic moment
The nuclear magnetic dipole moment is one of the primary observables by which the nuclear wave function can be probed.
The present study focuses on the measurement and implications of the magnetic moment of 
the excited state with $I^{\pi} = 3/2^-$ of $^{75}$Cu.  
Two low-lying isomeric states are known for $^{75}$Cu  
at the excitation energies of 61.7~keV and 66.2~keV~\cite{petrone}, 
as shown in the inset of Fig.~\ref{fig_cu-level}.
Their spin parities are expected to be either $1/2^-$ or $3/2^-$,
with the latter being inherited from the ground states of lighter isotopes up to $^{73}$Cu.
The magnetic moment of the $3/2^-$ isomeric state, in conjunction with that of the $5/2^-$ ground state~\cite{flanagan},
provides a stringent test of the theoretical description of Cu isotopes.  
The excitation energies are usually used as the first test of any such theory,
but they do not directly reflect the structure of the wave functions. 
By contrast, the moments are more directly related to the wave functions.
Thus, once this moment is verified, we can catch a glimpse of the properties of the wave function.

% need of spin alignment
Although the magnetic moments play an important role, those of the excited states in extremely neutron-rich nuclei, 
such as $^{75}$Cu, have not been measured to date.
This is because there has been no way to produce spin alignment (rank-two orientation) for such nuclei.
In order to produce high spin alignment in $^{75}$Cu,
we employed the recently introduced two-step reaction method~\cite{ichikawa},
utilizing the close relation between the angular momentum transferred to the fragment
and the direction of the removed momentum~\cite{asahi-pol} in the second projectile fragmentation (PF) reaction,
where high production yields for rare-isotope (RI) beams were ensured by combining
a technique of momentum-dispersion matching in the ion optics used for beam transport.
In Ref.~\cite{ichikawa}, the two-step PF scheme was demonstrated for the production of the excited state with $I^{\pi}=4^+$ of $^{32}$Al
from the projectile of $^{48}$Ca via one-neutron removal from an intermediate product of $^{33}$Al ($I^{\pi}=5/2^+$),
and 8\% spin alignment was achieved in spite of the mass difference between the projectile and the final fragment.
In the present work, the one-nucleon removal from the $I^{\pi}=0^+$ state was employed as the second PF reaction.
When a proton occupying an orbital with the angular momentum $j$ is removed from the $0^+$ state of $^{76}$Zn,
a state which spin $I(=j)$ should be preferentially populated in the final fragment of $^{75}$Cu,
and the relation between the direction of spin and the kinematics is considered to be refined,
thus, the spin alignment can be enhanced as compared to Ref.~\cite{ichikawa}.

% Experiment
The experiment was conducted using the in-flight superconducting RI separator BigRIPS~\cite{bigrips} at the RIKEN RIBF facility~\cite{ribf}
by means of the aforementioned two-step PF scheme.
The $^{75}$Cu beam was produced via a PF reaction for the removal of one proton from $^{76}$Zn, 
which was a product of the in-flight fission of a primary beam of $^{238}$U.  
The $^{75}$Cu beam was then introduced into an experimental apparatus at focal plane F8
for time-differential perturbed angular distribution (TDPAD) measurements.
(See the ``Methods'' section for the experimental details.)

\begin{figure}[tbp]
  \begin{center}
	\includegraphics[width=7.0cm]{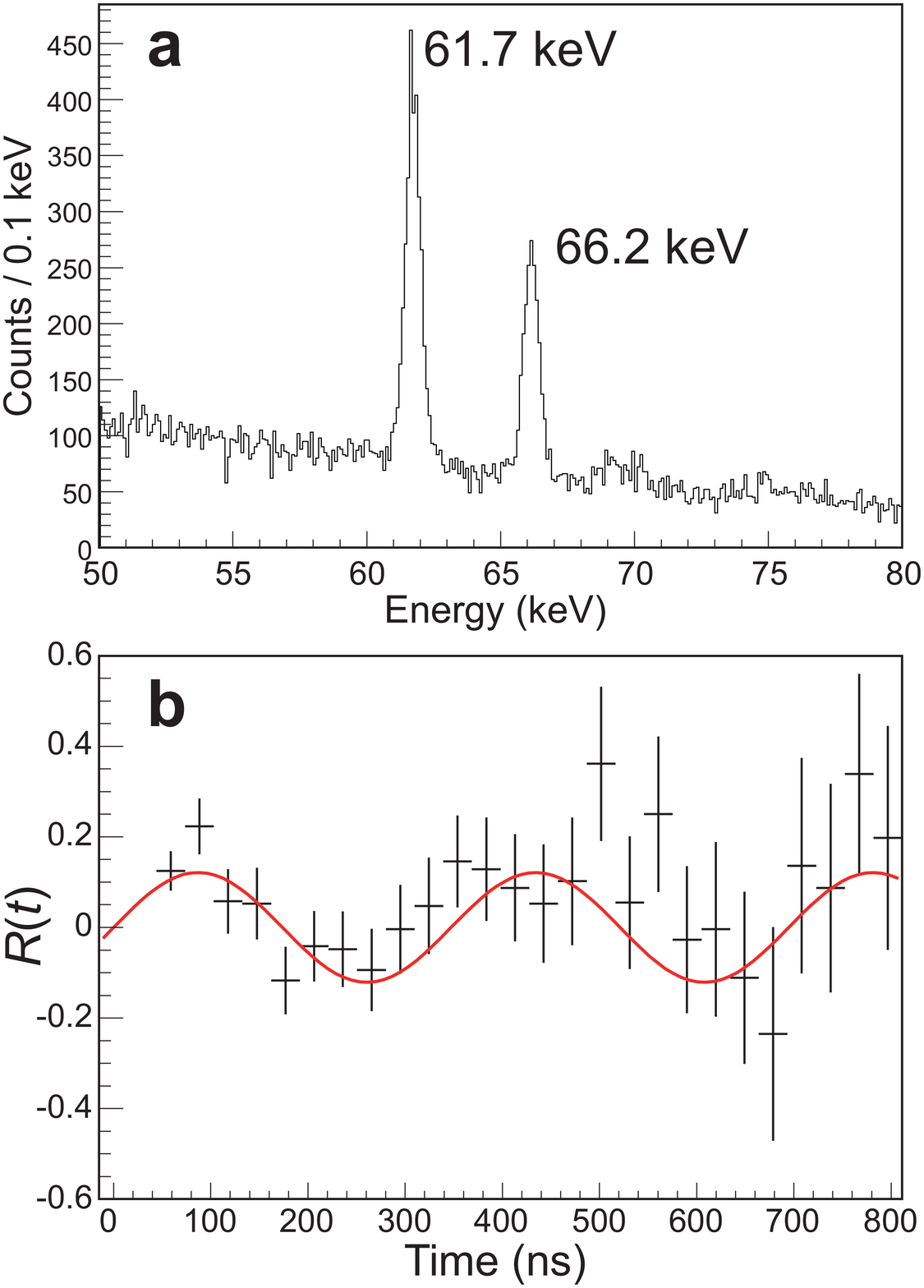}
  \end{center}
  \caption{Experimental results.
    {\bf a.} Observed $\gamma$-ray energy spectrum.
    The two peaks corresponding to the 61.7-keV and 66.2-keV $\gamma$ rays are labeled.
    {\bf b.} $R(t)$ ratios deduced from $N_{13}(t)$ and $N_{24}$(t), in accordance with Eq.~(\ref{eq_rfunc1}),
    for the 66.2-keV $\gamma$ rays.
    Error bars represent the standard deviation.
    The solid line represents the theoretical $R(t)$ function expressed in Eq.~(\ref{eq_rfunc2}) after
    fitting to the experimental $R(t)$ plot.
    See ``Methods'' for these equations.
}
	  {\label{fig_results}}
\end{figure}

% Results
Figure \ref{fig_results}~{\bf a} shows the $\gamma$-ray energy spectrum
where peaks corresponding to the 61.7-keV and 66.2-keV $\gamma$ rays are clearly observed.
$R(t)$ ratios representing the anisotropy for both $\gamma$ rays were evaluated (see "Methods")
and an oscillatory pattern was observed only for the 66.2-keV $\gamma$ ray, as shown in Fig.~\ref{fig_results}~{\bf b},
where the statistical significance of the oscillation was found to be greater than 5$\sigma$.
The spin parity of the 66.2-keV level was firmly identified as $3/2^-$ from the clear oscillation observed in the $R(t)$ ratio.
On the other hand, the spin parity of the 61.7-keV level, for which no oscillatory pattern was observed,
is most probably $1/2^-$,
because the rank-two anisotropy parameter ($A_{22}$ in Eq.~(\ref{eq_rfunc2}) in "Methods") is identically zero for the $I=1/2$ system,
inducing no oscillation in the TDPAD spectrum.
The $g$ factor of the 66.2-keV isomer in $^{75}$Cu was determined for the first time to be $g=0.93(4)$:
thus, the magnetic moment was found to be $\mu = 1.40(6) \mu_{\rm N}$ in units of the nuclear magneton $\mu_{\rm N}$.

% spin alignment
The magnitude of the spin alignment in the $^{75}$Cu was found to be 30(5)\%.
Although the statistical yield tends to be considerably small for exotic nuclei such as $^{75}$Cu,
5$\sigma$ significance is achieved in the TDPAD spectrum even with only 2,000 events of the $\gamma$-ray detection
owing to the high spin alignment.
Indeed, the 30(5)\% spin alignment realized shows only a slight reduction from a maximum value, 41\%, 
estimated in a similar way to Ref.~\cite{hufner} with a consideration of the momentum acceptance.
The alleviated reduction effect on the spin alignment by virtue of the refined choice of reaction to populate the $3/2^-$ state,
demonstrates an advantageous feature of the two-step PF scheme incorporating the $j-I$ correspondence.

\begin{figure}[tbp]
  \begin{center}
        \includegraphics[width=12cm]{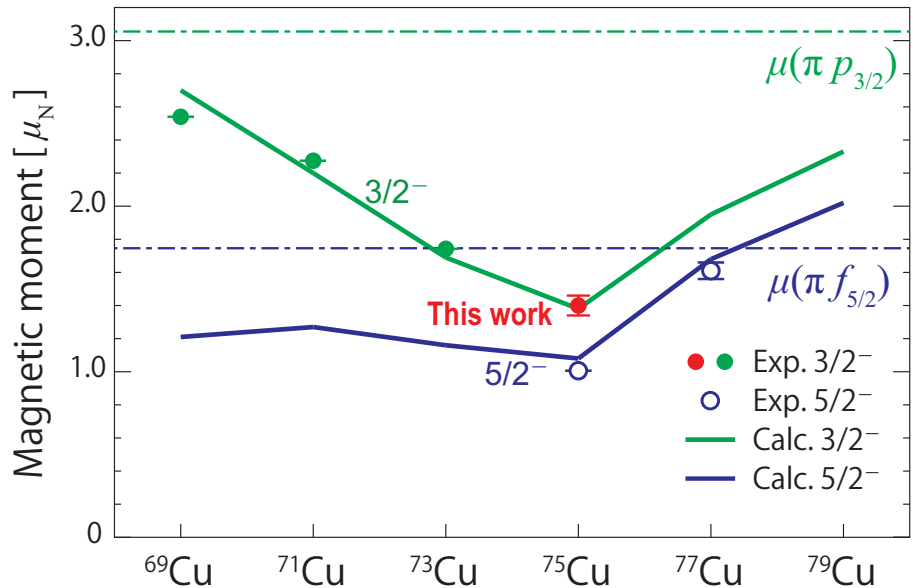}
  \end{center}
  \caption{Systematics of the magnetic moments for odd-$A$ Cu isotopes.
    Filled (open) circles represent experimental data for the $3/2^-$ ($5/2^-$) states~\cite{flanagan,koester,stone-71cu},
    with error bars of 1$\sigma$.
    The filled red circle represents the result obtained in this work.
    The solid green (blue) lines indicates the present MCSM calculations for the $3/2^-$ ($5/2^-$) states.
    $\mu (\pi p_{3/2})$ and $\mu (\pi f_{5/2})$ denote the proton Schmidt values for $p_{3/2}$ and $f_{5/2}$, respectively.
}
          {\label{fig_sys}}
\end{figure}

% general decomposition
The nuclear magnetic moment is evaluated usually in the nuclear structure calculation by the $\mbox{\boldmath $\mu$}$ operator~\cite{bohr-mottelson,Talmi},
\begin{eqnarray}
\label{eq_mu}
\mbox{\boldmath $\mu$} = \left\{ g_s({\rm p}) {\bf s}_{\rm p} + g_s({\rm n}) {\bf s}_{\rm n}
+ g_{\ell}(\rm p) \mbox{\boldmath $\ell$}_{\rm p} + g_{\ell}(\rm n) \mbox{\boldmath $\ell$}_{\rm n} \right\} \mu_{\rm N},
\end{eqnarray}
where ${\bf s}_{\rm p(n)}$ and $\mbox{\boldmath $\ell$}_{\rm p(n)}$ represent contributions from proton (neutron)
spin and orbital angular momentum, respectively.
The coefficients, $g_s$ and $g_{\ell}$ called spin and orbital $g$ factors, respectively,
carry not only free-nucleon values but also corrections such as the meson-exchange and the in-medium effects.
This work uses standard values, $g_s({\rm p})=3.91$, $g_s({\rm n})=-2.68$, $g_{\ell}({\rm p})=1.1$ and $g_{\ell}({\rm n})=-0.1$.
The present $I^{\pi} =3/2^-$ state can be interpreted, in the na\"ive picture, as a system consisting of a single proton in 
an appropriate orbital around an inert core.   
The magnetic moment in the single-particle limit (the Schmidt value) can then be calculated for a proton in the $p_{3/2}$ orbital,
resulting in a value of $\mu(\pi p_{3/2}) = 3.05 \mu_{\rm N}$.
Figure~\ref{fig_sys} depicts the measured and calculated magnetic moments of $^{69-79}$Cu.
The measured value for $^{69}$Cu appears to be not too far from the Schmidt value,
suggesting that the picture of a single proton orbiting an inert core makes sense.
However, Fig.~\ref{fig_sys} indicates that the measured value deviates greatly as $N$ increases.

% MCSM calculation
To investigate the implications of this growing deviation, theoretical studies were conducted using
the Monte Carlo Shell Model (MCSM)~\cite{shimizu,otsuka2001}.
The same Hamiltonian (A3DA-m) and the same large model space, $20 \leq N(Z) \leq 56$,
that have previously been used in successful calculations for Ni isotopes~\cite{tsunoda} were employed.
The magnetic moments were calculated from wave functions thus obtained with the $g$ factors in Eq.~(\ref{eq_mu}).
The results shown in Fig.~\ref{fig_sys} demonstrate a remarkably consistent trend compared to the experimental values
of all measured magnetic moments of the $5/2^-$ and $3/2^-$ states,
including the deviation from the Schmidt value which appears to be maximal at $^{75}$Cu.

% preserve or destroy
By taking the deviation from the Schmidt value, one can extract the effect of the core excitations.  
It is thus found that this effect is rather sizable, but the calculated magnetic moments still agree well with the experimental values.
Let us now investigate whether this effect destroys the shell evolution picture.

\begin{figure*}[tbp]
  \begin{center}
	\includegraphics[width=15cm]{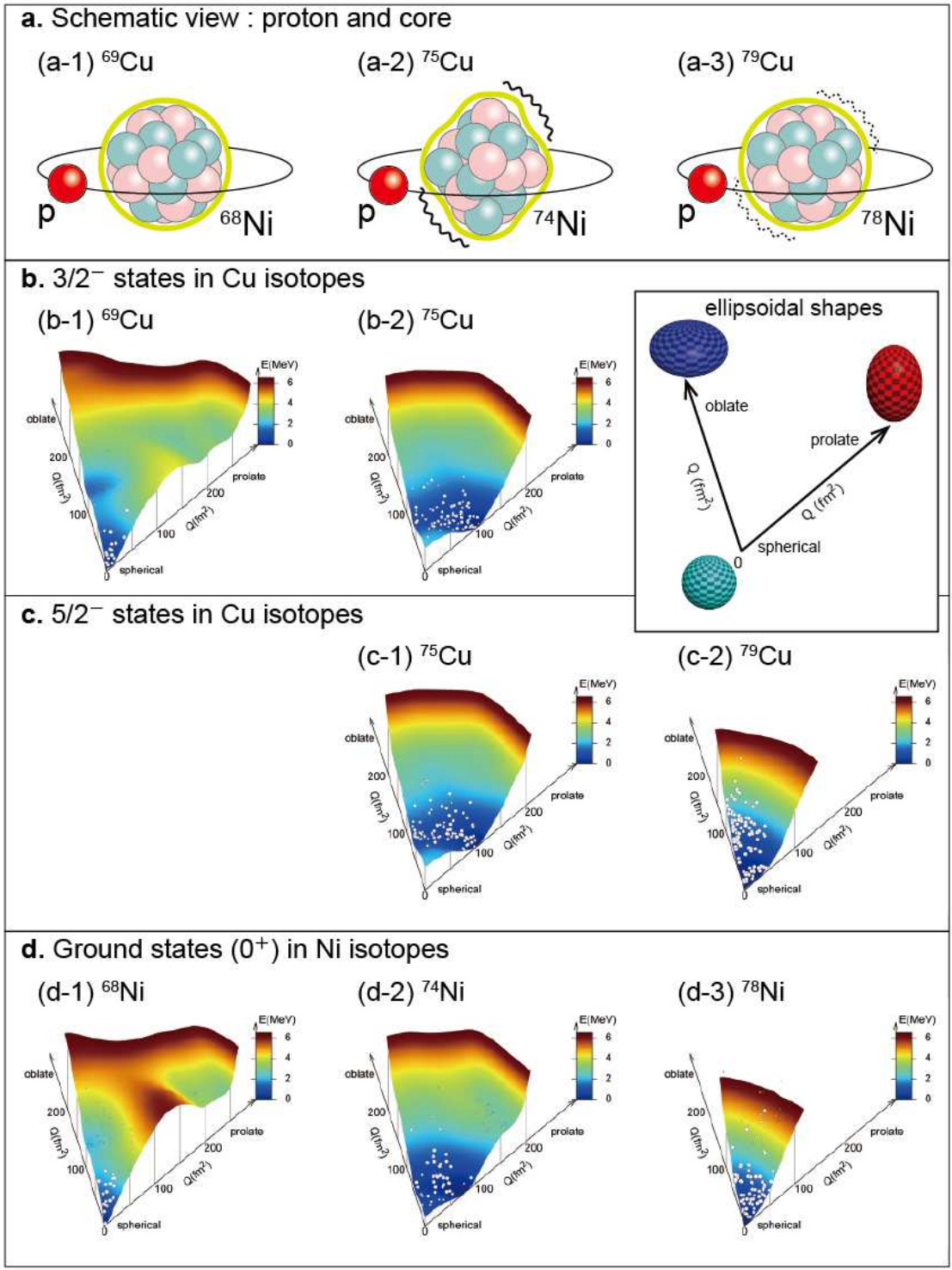}
  \end{center}
  \caption{Theoretical interpretations.
    {\bf a.} Illustrations of a Cu isotope as a proton orbiting around (a-1) a spherical Ni core, (a-2) a Ni core with 
    excitations (surface oscillations) and (a-3) a Ni core with weak excitations.      
    T-plots for {\bf b.} the $3/2^-$ states in the Cu isotopes (b-1) $^{69}$Cu and (b-2) $^{75}$Cu, 
    {\bf c.} the $5/2^-$ states in the Cu isotopes (c-1) $^{75}$Cu and  (c-2) $^{79}$Cu,
    and {\bf d.} the ground states in the Ni isotopes (d-1) $^{68}$Ni, (d-2) $^{74}$Ni and (d-3) $^{78}$Ni,
    where the potential energy surface relative to the minimum is represented by a 3D contour with two independent  
    coordinates of quadrupole moments, as drawn in the inset.    
  }
	  {\label{fig_tplot}}
\end{figure*}

% T-plot systematics
Figure~\ref{fig_tplot}~{\bf b}, {\bf c} and {\bf d} show T-plots, which visualize major components of MCSM wave functions by white circles 
(see the ``Methods'' section)~\cite{tsunoda,otsuka2016}. 
For $N=40$, being a magic number, the T-plots for $^{68}$Ni and $^{69}$Cu, shown in Fig.~\ref{fig_tplot}~(d-1) and (b-1), respectively,
show many similarities, including a concentration around a spherical shape.      
Thus, the $3/2^-$ state in $^{69}$Cu is considered to be a proton in the $p_{3/2}$ orbital on top of a spherical core of $^{68}$Ni,
as illustrated in Fig.~\ref{fig_tplot}~(a-1).
Figures~\ref{fig_tplot} (b-2) and (c-1) show those for the $3/2^-$ and $5/2^-$ states of $^{75}$Cu, respectively,
where the T-plots are spread more widely than the $N=40$ case because the excess neutrons produce more shape oscillations.  
This characteristic feature is further shared by $^{74}$Ni shown in Fig.~\ref{fig_tplot}~(d-2).    
Thus, the $3/2^-$ and $5/2^-$ states of $^{75}$Cu are considered to correspond, to a large extent, 
to a proton orbiting around a $^{74}$Ni core that exhibits certain excitations or shape oscillations
from a perfect sphere, as illustrated intuitively in Fig.~\ref{fig_tplot}~(a-2).  

% shell evolution and collectivity
Now that this background has been elucidated, we shall return to the interplay between the shell evolution and the collective mode,
or the core excitation.         
The effect of the core excitation varies as $N$ increases, as shown in Fig.~\ref{fig_sys},
mainly due to the growing occupancy of neutrons in the $g_{9/2}$ orbital,
and it differs between the $3/2^-$ and $5/2^-$ states because of alternating couplings.
The deviation from the Schmidt value is maximal at $^{75}$Cu,
however the picture of a proton orbiting around a correlated Ni core still appears to be reasonable for this nucleus.

%energy
Coming back to Fig.~\ref{fig_cu-level}~{\bf b},
the MCSM full calculation shows distinct decrease in energy up until the inversion near $^{75}$Cu.  
The slope of the ``shell evolution'' curve resembles that of the full calculation, and the energy gain from the former to the latter
is mainly a result of core-excitation effects (the vertical arrow).  
We also calculated the energy expectation values for the same wave functions by setting equal
the $\pi f_{5/2}$-$\nu g_{9/2}$ and $\pi p_{3/2}$-$\nu g_{9/2}$ monopole interactions,
thereby eliminating most of the shell-evolution effects.
This is labelled as ``core excitation'', which is remarkably close to constant with no symptom of inversion.
Thus, the shell evolution indeed arises even in the presence of core excitations,
whereas the core excitation is not a major driving force in the present case.
The nearly constant energy shift due to the core excitation moves the inversion point on the $N$ axis to the left
(the horizontal arrow).
This feature reinforces the concept of shell evolution rather than destroying it.
Present and past measurements of the magnetic moment verify such structural evolution.
In particular, the simultaneous agreement for the two states in $^{75}$Cu is crucial,
leading to the conclusion that the shell evolves even in the presence of the collective modes.
(See "Methods" for related uncertainties).

% outlook towards 78Ni
The present structural evolution scenario yields an intriguing prediction of the closed-shell properties
of the doubly magic nucleus $^{78}$Ni and its neighbour $^{79}$Cu~\cite{olivier,welker}.
Their T-plots, shown in Fig.~\ref{fig_tplot}~(c-2) and (d-3), depict the profound minima of
the potential energy surface at the spherical shape.
T-plot points are clustered around the spherical shape in both $^{78}$Ni and $^{79}$Cu,
however, more spreading or fluctuation is evident than in $^{68}$Ni or $^{69}$Cu.
This is because the magic number of $N=40$ lies between major shells with opposite parities,
whereas the magic number of $N=50$ is due to spin-orbit splitting, and excitation can occur by one particle.
The present study thus foresees the existence of
a spherical doubly magic $^{78}$Ni nucleus, while fluctuations may be stronger.
Therefore, $^{78}$Ni can be the first {\it jj}-coupled doubly magic nucleus with significant sphericity~\cite{hagen,nowacki}.
The implications of such properties for the nucleosynthesis are of considerable interest, for instance, through $\beta$-decay rates~\cite{astro}.

\section*{Methods}
{\bf Two-step projectile fragmentation scheme:}
In the present experiment the two-step PF scheme was employed in the production of spin-aligned $^{75}$Cu beams.
In the reaction at the primary target position F0, $^{76}$Zn (with one proton added to $^{75}$Cu)
was produced by the in-flight fission of a 345-MeV/nucleon $^{238}$U beam on a $^9$Be target
of 1.29 g/cm$^2$ in thickness.
This thickness was chosen to provide the maximal production yield for the secondary $^{76}$Zn beam.
A wedge-shaped aluminium degrader with a mean thickness of 1.65~g/cm$^2$ was placed
at the first momentum-dispersive focal plane F1, where the momentum acceptance was $\pm$3\%.
The secondary $^{76}$Zn beam was transmitted to a second wedge-shaped aluminum plate
with a mean thickness of 0.81~g/cm$^2$ placed at the second momentum-dispersive focal plane F5.
Thus, $^{75}$Cu nuclei (including those in the isomeric state $^{75{\rm m}}$Cu) were produced
through a PF reaction involving the removal of one neutron from $^{76}$Zn.
The target had a thickness of 3~mm: consequently, the spread of the momentum due to the dispersal of the reaction positions
in the secondary target is comparable to the width of the momentum distribution~\cite{goldhaber},
$\sigma_{\rm G}=90$~MeV/$c$ for one-nucleon removal.
The $^{75}$Cu beam was subsequently transported to the double-achromatic focal plane F7
under conditions such that the momentum dispersion generated at the site of the first reaction was effectively cancelled out.
The momentum selection to produce the spin alignment was performed using information obtained
from two parallel-plate avalanche counters placed at F7.
Only events in which beam particles were detected within a horizontal region of $\pm 12$~mm were recorded and analysed offline.

\vspace{0.4cm}
% TDPAD
{\bf Time-differential perturbed angular distribution (TDPAD) measurement:}
The TDPAD apparatus consisted of an annealed Cu metal stopper, a dipole magnet, Ge detectors,
a plastic scintillator and a collimator.
The Cu stopper was 3.0~mm in thickness and 30$\times$30~mm$^2$ in area,
and the dipole magnet provided a static magnetic field of $B_0=0.200$~T.
De-excitation $\gamma$ rays were detected by four Ge detectors 
perpendicular to the magnetic field
at angles of $\pm45^{\circ}$ and $\pm135^{\circ}$ with respect to the beam axis.
Three of the four Ge detectors were of the planar type (Low-Energy Photon Spectrometer) and
the remaining one was of the coaxial type.
The plastic scintillator, which was of 0.1~mm in thickness, was placed upstream of the stopper, 
and the signal from the scintillator served as a time zero for the TDPAD spectrum.

TDPAD measurements enable us to determine the $g$ factor by observing the time-dependent anisotropy of
the de-excitation $\gamma$ rays emitted from nuclei during spin precession under the external magnetic field.
From the experimental findings, a ratio $R(t)$, defined as
\begin{equation}
  {\label{eq_rfunc1}}
  R(t) = \frac{N_{13}(t) - \epsilon N_{24}(t)}{N_{13}(t) + \epsilon N_{24}(t)},
\end{equation}
is evaluated, where $N_{13}(t)$ and $N_{24}(t)$ are the respective sums of the photo-peak counting rates
at two pairs of Ge detectors placed diagonally to each other
and $\epsilon$ is a correction factor for the detection efficiency.
Theoretically, $R(t)$ is also given by 
\begin{equation}
  {\label{eq_rfunc2}}
 R(t) = -\frac{3 A_{22}}{4 + A_{22}} \sin(2\omega_{\rm L} t).
\end{equation}
Here, the Larmor frequency $\omega_{\rm L}$ is given by 
$ \omega_{\rm L} = g \mu_{\rm N} B_0 / \hbar$,
where $g$ is the $g$ factor for $^{75}$Cu, and has a relation to the magnetic moment $\mu$ and the nuclear spin $I$,
as $\mu = g I \mu_{\rm N}$. 
The parameter $A_{22}$ is the rank-two anisotropy parameter,
defined as $A_{22}=AB_2F_2$,
where $A$ denotes the degree of spin alignment, $B_2$ is the statistical tensor for complete alignment,
and $F_2$ is the radiation parameter~\cite{morinaga}.

In the present experiment, the $R(t)$ ratio was evaluated in accordance with Eq.~(\ref{eq_rfunc1}).
The statistical significance of the oscillation for the 66.2-keV $\gamma$ ray was
found to be greater than 5$\sigma$, with $A_{22}=-0.17 (3)$,
which was obtained from a least-$\chi ^2$ fit of the theoretical function
given in Eq.~(\ref{eq_rfunc2}) to the experimental one by Eq.~(\ref{eq_rfunc1}).

\vspace{0.4cm}
% MCSM
{\bf Monte Carlo Shell Model (MCSM):}
Shell-model calculations in nuclear physics are quite similar to configuration interaction (CI) 
calculations in other fields of science.
The major differences are that (i) the two ingredients are protons and neutrons instead of electrons
and (ii) nuclear forces are considered instead of Coulomb or other forces.
Conventionally, the matrix of the Hamiltonian with respect to many Slater determinants is diagonalized.   
Because many configurations are needed in large systems, the dimensions of the matrix can be enormous,
making the calculation infeasible: however, many interesting and important problems lie beyond this limit.
MCSM represents a breakthrough in this regard.  
A set of Slater determinants, called MCSM basis vectors, is introduced,
and the diagonalization is performed in the Hilbert space spanned by the MCSM basis vectors.  
Each MCSM basis vector is a Slater determinant composed of single-particle states
that are superpositions of the original single-particle states,
and the amplitudes of these superpositions are determined through a combination of stochastic and variational methods.
Even when the dimensions are on the order of 10$^{23}$ in the conventional shell model, the problem can be solved,
to a good approximation, with up to approximately 100 MCSM basis vectors.
This method has seen many applications up to Zr isotopes.

% shell model interaction
The present MCSM calculations were performed with the A3DA-m interaction.
This interaction was fixed, prior to this work, based on the microscopic G-matrix with certain empirical modifications
so as to describe experimental energy levels of Ni isotopes~\cite{tsunoda}. 
The validity of this interaction has been confirmed independently in recent studies on different topics~\cite{groote,wraith}.
The MCSM calculation with such a large model space employed in this work can describe both the excited state and the ground state appropriately,
involving various cross-shell excitations.

% uncertainty
The magnetic moment is calculated by Eq.~\ref{eq_mu} with four coefficients, $g_s({\rm p, n})$ and $g_{\ell}({\rm p, n})$,
for which some uncertainties exist.
Following the usual prescription~\cite{Castel-Towner,Towner1987},
the spin $g$ factors $g_s({\rm p})$ and $g_s({\rm n})$ are reduced from their free-space values,
represented by the so-called spin quenching factor $q_s$.
The empirically obtained appropriate value of $q_s$ is in or around the range from 0.70 to 0.75 for the calculations
in an incomplete $LS$ (or Harmonic Oscillator) shell~\cite{jun45,rmp_caurier}, to which the present work belongs.
The value 0.70 is used in the main text as in Ref.~\cite{jun45}.
The orbital $g$ factors $g_{\ell}({\rm p})$ and $g_{\ell}({\rm n})$ take values, 
$g^{\rm free}_{\ell}({\rm p})=1$ and $g^{\rm free}_{\ell}({\rm n})=0$, in the free space, respectively.
In nuclear medium, the isovector correction based on the meson-exchange picture is made usually:
$g_{\ell}({\rm p}) = g_{\ell}^{\rm free}({\rm p})  + \delta g_{\ell}$ and 
$g_{\ell}({\rm n}) = g_{\ell}^{\rm free}({\rm n}) - \delta g_{\ell}$~\cite{Castel-Towner,Towner1987},
with the empirical value of $\delta g_{\ell}$ being between 0~\cite{jun45} and 0.1~\cite{gxpf1}.

The measured magnetic moments of $^{75}$Cu are $1.40(6) \mu_{\rm N}$ for the first $3/2^-$ state as obtained by the present work,
and $1.0062(13) \mu_{\rm N}$ for the first $5/2^-$ state~\cite{flanagan}.
We can evaluate the values of $q_s$ and $\delta g_{\ell}$ from them.
Because of the small error of the $5/2^-$-state moment, it is treated as a single value, $1.0062$.
By moving within the error bar of the $3/2^-$-state moment, the values of $q_s$ and $\delta g_{\ell}$ move from 
$(q_s, \delta g_{\ell})=(0.67,\sim 0)$ to $(0.78, 0.13)$.
This range is essentially within the uncertainties of $q_s$ and $\delta g_{\ell}$.
The two measured magnetic moments can thus be nicely reproduced with the present eigen wave functions of
the first $3/2^-$ and $5/2^-$ states, and the main conclusion is insensitive to the uncertainties associated with $q_s$ and $\delta g_{\ell}$.
We note that the simultaneous determination of the values of $q_s$ and $\delta g_{\ell}$ can provide a precious insight as shown here,
spotlighting the advantage of the measurement of more than one magnetic moment for a single nucleus.

We can further investigate another uncertainty coming from the shell-evolution strength.
In Fig.~\ref{fig_cu-level}~{\bf b}, a certain relevant monopole interaction is removed to obtain the "core excitation" result.
Instead of removing fully, we now weaken this particular monopole interaction, and see how the magnetic moments are changed.
The actual MCSM calculations indicate that if the 10\% weakening (90\% remaining) of this monopole interaction is made, 
the $(q_s, \delta g_{\ell})$ values obtained similarly appear to be between $(0.62, -0.05)$ and $(0.71, 0.07)$.
The situation is shifted to or beyond the edge of the uncertainties of $q_s$ and $\delta g_{\ell}$, and a further reduction will kick us out almost completely.
The uncertainty of the monopole interaction strength is now deduced, based on the magnetic moments, to be about 10\% of its strength.
Such uncertainty is so small that the shell evolution is quite robust in the present case.

\vspace{0.4cm}
{\bf T-plots:}
%T plot
A T-plot is a method of analysing an MCSM wave function.
The intrinsic quadrupole moments are calculated for each MCSM basis vector, and the corresponding vector is identified
by a circle on the potential energy surface with those quadrupole moments as coordinates.  
The overlap probability of each MCSM basis vector with the eigenstate being considered 
is represented by the size of the circle as an indicator of its importance. 
Angular momentum/parity projection is performed in this process.      
Thus, with a T-plot, one can visualize the shape characteristics of an MCSM basis vector and its importance for a given eigenstate.
The main area indicates the shape and the spreading represents the extent of quantum fluctuations.

\vspace{0.4cm}
{\bf Detailed theoretical outputs:}
% occupation number
The MCSM calculation indicates that in the $3/2^-$ state of $^{75}$Cu
the occupation number of the proton $p_{3/2}$ orbital is 0.86, somewhat smaller than unity,
whereas that of the proton $f_{5/2}$ orbital is 0.35.  
Likewise, in the $5/2^-$ state, the orbital $p_{3/2}$ is occupied with a similarly small probability.
Thus, single-particle nature remains to a certain extent. 

% spectroscopic factor
This argument of the T-plot for $^{75}$Cu can be theoretically quantified in terms of the so-called spectroscopic factor, 
0.50 for $\pi p_{3/2}$ and 0.46 for $\pi f_{5/2}$, from $^{74}$Ni to $^{75}$Cu. 
These values represent the probabilities that the $I^{\pi} =3/2^-$ and $5/2^-$ states of $^{75}$Cu are
created by adding one proton to the ground state of $^{74}$Ni without any disturbance.
Although these probabilities are rather large, other states of the core,
such as various surface oscillations, must account for most of the remaining probability.
Figure~\ref{fig_tplot}~(a-2) schematically illustrates such a situation of
single-particle motion on top of a core exhibiting various excitations.

\vspace{0.4cm}
{\bf Behavior of the $1/2^-$ state:}
% 1/2- state
In the present experiment the spin parity of the 61.7-keV level was assigned to be $I^{\pi}=1/2^-$ based on the non-observation of the oscillation signal.
At a glance, the level energies of the $1/2^-$ (including $(1/2^-)$) states of the Cu isotopes
seems to be connected with the lowering of the $5/2^-$ states, as shown in Fig.~\ref{fig_cu-level}~(a).
The nature of the $1/2^-$ state of $^{75}$Cu is discussed in terms of the transition probability which was firstly measured
and discussed for assumed level schemes in Ref.~\cite{petrone}.
Based on the level scheme fixed by the present study, the reduced transition probability from the $1/2^-$ state to the $5/2^-$ ground state was determined to be 
$B({\rm E2})=22.5(8)$ in Weisskopf units which was well reproduced by the MCSM output $B({\rm E2})=23.5$ in Weisskopf units.
Referring to the wave function by the MCSM output with a notice that the magnetic moment of the $5/2^-$ state is well reproduced,
it could be supposed that the core of the $1/2^-$ state is rather similar to those of the $3/2^-$ and $5/2^-$ states,
whereas the valence proton occupies the $f_{5/2}$, $p_{3/2}$ and $p_{1/2}$ orbitals with considerable admixtures.
As for $^{69,71,73}$Cu, the $B({\rm E2})$ values from the $1/2^-$ states (including those in brackets)
to the $3/2-$ ground state have been reported~\cite{stefanescu},
and an increasing trend of $B({\rm E2})$ over $N=40$, combined with the lowering of the level energies of the $1/2^-$ states,
might indicate rather collective nature of those states.

\section*{Data availability}
The data used in the present study are available from the corresponding author upon reasonable request.

\begin{acknowledgments}
The experiment was performed under Program No.~NP1412-RIBF124R1 at RIBF,
operated by RIKEN Nishina Center for Accelerator-Based Science and CNS, The University of Tokyo. 
We thank the RIKEN accelerator staff for their cooperation during the experiment.
This work was supported in part by a JSPS KAKENHI grant (16K05390),
and also in part by JSPS and CNRS under the Japan-France Research Cooperative Program.
The MCSM calculations were performed on the K computer at RIKEN AICS (hp140210, hp150224, hp160211, hp170230).
This work was also supported in part by the HPCI Strategic Program (The origin of matter and the universe)
and ``Priority Issue on Post-K computer'' (Elucidation of the Fundamental Laws and Evolution of the Universe)
D.L.B. and A.K. acknowledge support by the Extreme Light Infrastructure Nuclear Physics (ELI-NP) Phase II,
a project co-financed by the Romanian Government and the European Union through the European Regional
Development Fund -- the Competitiveness Operational Programme(1/07.07.2016, COP, ID 1334).
D.R. was supported by the P2IO excellence center.
%Japan-French
\end{acknowledgments}

\section*{Author contributions}
Y.I. designed the experiment, analysed the data and chiefly wrote the paper.
Y.T. and T.O. worked on the theoretical studies, and T.O. partly wrote the paper.
The other authors are collaborators on the experiment.

%\section*{Competing financial interest}
%The authors declare no competing financial interests.

\end{document}